\documentclass{aa}  

\usepackage[hidelinks]{hyperref}
\usepackage{graphicx}
\usepackage{txfonts}
\usepackage{lineno}
\usepackage{soul}

\usepackage{comment}
\usepackage[printonlyused]{acronym}
\usepackage{mathtools}
\usepackage{stmaryrd}
\DeclarePairedDelimiter\abs{\lvert}{\rvert}%
\DeclarePairedDelimiter\norm{\lVert}{\rVert}%
\makeatletter
\let\oldabs\abs
\def\abs{\@ifstar{\oldabs}{\oldabs*}}
\let\oldnorm\norm
\def\norm{\@ifstar{\oldnorm}{\oldnorm*}}
\makeatother

\usepackage{amsfonts}
\usepackage{amsmath}

\usepackage{subfig}
\usepackage{xcolor} % Include the xcolor package

\usepackage{fancyhdr}
\renewcommand{\and}{   {,\ }}
\newcommand{\affil}{   {}}

\begin{document}

   \title{Multimodal Flare Forecasting with Deep Learning}

   % \subtitle{}

\author{
   G. Francisco$^{\affil{1,2,3}}$, \
   S. Guastavino$^{\affil{4}}$, \
   J. Fernandes$^{\affil{5}}$, \
   T. Barata$^{\affil{2}}$ and \
   D. Del Moro$^{\affil{1}}$
}
% \author{
%    G. Francisco\affil{1,2,3} 
%    S. Guastavino\affil{4} 
%    J. Fernandes\affil{5}
%    T. Barata\affil{2}
%    D. Del Moro\affil{1}
% }

% \author[1,2,3]{G. Francisco}
% \author[4]{S. Guastavino}
% \author[5]{J. Fernandes}
% \author[2]{T. Barata}
% \author[1]{D. Del Moro}

   \institute{Department of Physics, University of Rome Tor Vergata, Rome, Italy\\
              \email{gregoire.francisco@gmail.com}
         \and
             IA, Instituto De Astrofisica E Ciências Do Espaço, University of Coimbra, Coimbra, Portugal%\\
         \and
             Department of Physics, University of Rome La Sapienza, Rome, Italy%\\
        \and
         MIDA, Dipartimento di Matematica, Universita` di Genova, Genova, Italy%\\
         \and
         CITEUC, Geophysical and Astronomical Observatory, University of Coimbra and Department of Mathematics, Coimbra, Portugal
         }

% \author{
%    G. Francisco$^{1,2,3}$,
%    S. Guastavino$^{4}$,
%    J. Fernandes$^{5}$,
%    T. Barata$^{2}$,
%    D. Del Moro$^{1}$
% }

% \affil[1]{Department of Physics, University of Rome Tor Vergata, Rome, Italy}
% \affil[2]{IA, Instituto De Astrofísica e Ciências Do Espaço, University of Coimbra, Coimbra, Portugal}
% \affil[3]{Department of Physics, University of Rome La Sapienza, Rome, Italy}
% \affil[4]{MIDA, Dipartimento di Matematica, Università di Genova, Genova, Italy}
% \affil[5]{CITEUC, Geophysical and Astronomical Observatory, University of Coimbra and Department of Mathematics, Coimbra, Portugal}
% \affil[ ]{\texttt{gregoire.francisco@gmail.com}}

   % \date{Received September 15, 1996; accepted March 16, 1997}

\abstract{Solar flare forecasting mainly relies on photospheric magnetograms and associated physical features to predict forthcoming flares. 
   However, it is believed that flare initiation mechanisms often originate in the chromosphere and the lower corona. 
   In this study, we employ deep learning as a purely data-driven approach to compare the predictive capabilities of chromospheric and coronal UV and EUV emissions across different wavelengths with those of photospheric line-of-sight magnetograms. 
   Our findings indicate that individual EUV wavelengths can provide discriminatory power comparable  or better to that of line-of-sight magnetograms. Moreover, we identify simple multimodal neural network architectures that consistently outperform single-input models, showing complementarity between the flare precursors that can be extracted from the distinct layers of the solar atmosphere. 
   To mitigate potential biases from known misattributions in Active Region flare catalogs, our models are trained and evaluated using full-disk images and a comprehensive flare event catalog at the full-disk level. 
   We introduce a deep-learning architecture suited for extracting temporal features from full-disk videos.
   \vspace{10ex}
   }%{}{}{}{} 
   
\maketitle

% 5 {} token are mandatory
% \maketitle \footnote{Part of this manuscript is based on a preprint previously posted on ESS Open Archive (10.22541/essoar.170688972.24631782/v3) on May 15, 2024. As the original work was found too long, the current manuscript focus and expand on the comparative study of distinct and multimodal SDO inputs for flare forecasting and add a video model (VideoLENS) to include temporal dynamic in the analysis. The preprint's original metric discussion and PCNN model are left out for another futur publication.}

% \abstract 


\keywords{Solar Flares --
         Deep Learning --
         Corona --
         Chromosphere --
         Attention 
           }

   % \maketitle
%
%-------------------------------------------------------------------
\section{Introduction}

Solar flares are sudden bursts of electromagnetic radiation and energetic particles that can pose significant threats to human health and technology. 
Their potential danger is typically assessed based on the intensity of the emitted soft \ac{SXR} flux. 
Flares are classified into five categories: A, B, C, M, and X, with each class representing a peak SXR flux that is one order of magnitude higher than the previous class. 
The M-Class threshold marks the point from which flares are considered strong, while the X-Class threshold indicates flares that present the most serious threats to society.
A significant research effort is given on identifying physical precursors of flares, often derived from photospheric magnetograms. 
For example, the FLARECAST project identified 209 potential flare precursors, with 94\% of them related to \ac{AR} magnetic field properties (\cite{Georgoulis2021})
Another important trend, initiated by \cite{Huang2018}, involves using deep learning as a purely data-driven approach, employing neural networks to magnetogram images. 
In contrast, the potential of chromospheric and coronal observations for flare prediction remains less explored despite the belief that flare-triggering mechanisms originate in these regions \cite[][]{shibata2011solar, toriumi2019flare}. % TODO : add SHREEYESH when available
One barrier to investigating the upper atmospheric layers for flare precursors is the current observational limitations in deriving magnetograms and related physical quantities from these regions.
However, this challenge may soon be addressed at the chromospheric level with upcoming projects like the Solar Activity Monitor Network (SAMNet) \cite{Samnet} and the Global Automatic Telescopes for Exploring the Sun (GATES) \cite{giovannelli2020tor, konow2024gates} which aim to provide high temporal coverage of chromospheric magnetograms. 
Moreover, instruments like the \ac{SDO} (\cite{Chamberlin2012}) and its \ac{AIA} (\cite{Lemen2012}) already provide rich spectral data that may offer valuable thermal and morphological plasma features for both the chromosphere and the corona. 
In this context, \cite{Dissauer2023} introduced a dataset of \ac{AR} \ac{AIA} patches, from which \cite{Leka2023} derived chromospheric and coronal features based on moment analysis, achieving promising results. 
Notable precursors identified include total emission, steep brightness variations, and high-order moments of running differences, which suggest a tendency for short, small-scale brightening events in flare-imminent regions. 
Additionally, \cite{Sun2023} utilized \ac{CNN}s with \ac{AIA} coronal images to forecast flares above the M-class threshold, achieving state-of-the-art results.
Both \cite{Leka2023} and \cite{Sun2023} identified the 94\AA 
 line emission as particularly predictive of upcoming flares. 
\cite{Sun2023} further demonstrated that averaging predictions of single-wavelength models outperforms individual models' predictions.
The performance improvement achieved by ensembling models, such as model bagging, can typically be attributed to the law of large numbers.
Specifically, under the assumption of independent and identically distributed Gaussian errors, averaging models' outputs reduces error variance, as individual errors tend to cancel each other out. 
However, this method does not exploit potential dependencies between the distinct models' features, which might also enhance predictions.

In this study, we aim to extend the works of \cite{Leka2023} and \cite{Sun2023} by:
\begin{enumerate}
  \item Comparing the discriminative power of chromospheric and coronal features with those derived from line-of-sight magnetograms.
  \item Investigating potential linear and non-linear complementarities between features from different atmospheric layers using multimodal models.
  \item Exploring the discriminative power of long-term temporal features in addition to single-timestep features.
\end{enumerate}

We utilize Deep Learning as a purely data-driven approach on full-disk images produced by the \ac{SDO}. 
This full-disk approach complements the AR-level approach of \cite{Leka2023} and \cite{Sun2023}, and reduces mislabels due to known misattributions in \ac{AR} datasets. 
Such misattributions, as highlighted by \cite{Kiera2022}, can affect up to 8\% of flares above the M-class threshold, and up to 20\% of these events are simply missing in the standard GOES Flare catalog. 
Although the Heliophysics Event Knowledgebase (HEK) corrects many of these issues, some flares remain misattributed or unaccounted for, potentially compromising model training and evaluation in the already challenging context of imbalanced and scarce data.
% calculation from Kiera2022
% 20 misabeled MX (5. conclusion, p.10)
% 64 missing  (5. conclusion, p.10)
% 243 M + 16 X + 64 missing = 323 MX events p.4
% 20/ 243 = 8% misatribtuion among atributed ; 64 / 323 =  200%
Our focus is on binary forecasts of the occurrence of at least one flare above the M-class threshold within 24-hour windows. 
This paper is organized as follows: Section 2 describes the data, Section 3 describes the models and training methods, Section 4 describes the results and discussions.

\section{Data}

We use the SDO-2H-ML image dataset introduced in \cite{Francisco2024}.
This is a 2 hour resolution datasett ranging from May 2010 to April 2023, for a total amount of about 54000 dates for which are associated the corresponding \ac{LOS}-magnetogram and AIA images in the wavelength 1600\AA, 304\AA, 171\AA, 193\AA, 211\AA \ and 94\AA. 
The AIA images are preprocessed through alignment, exposure normalization, correction for instrument degradation, and are then downsampled to a resolution of 1024x1024 pixels. AIA and HMI pixel values undergo logarithmic scaling and are reduced to 8-bit depth, preserving the majority of the original pixel distribution with minimal information loss, while also reducing the dataset's size. The image are then compressed as JPEGs, resulting in another moderate loss of small-scale and high frequency details.
In an ablation study, we found in \cite{Francisco2024}, that such  JPEG compression, as well as a further downsampling to 448x448 pixels, had no significant impact on the resulting model performances. We therefore work again with the 448x448 resolution.
We use the same flare catalog as in  \cite{Francisco2024}, which is an extension of \cite{Plutino2023}'s catalog.
%--------------------------------------------------------------------
\section{Methodology}
%-------------------

\subsection{Models}

\subsubsection{Single Timestep Models}

% \textit{Architecture choice}
% \newline
% \newline
To predict flares using single-wavelength images or magnetograms, we employ transfer learning from models pre-trained on the ImageNet dataset \cite{ImageNet2009}. 
The extensive sample and class diversity in ImageNet enable models to learn general hierarchical filters and features that transfer well to different problems and datasets. 
Research by \cite{Kornblith2019} indicates that for new problems with limited data, fine-tuning all layers of a pre-trained model on the new dataset yields optimal performance, typically proportional to the model's original performance on ImageNet.
Consequently, we use the EfficientNetV2-S \cite{Tan2021efficientnetv2}. 
As of now, EfficientNetV2 models are among the best-performing CNN architectures, significantly outperforming ResNet models \cite{Resnet2015}. EfficientNetV2 incorporates advanced residual blocks, including Mobile Inverted Bottleneck Convolution (MBCConv) \cite{MobileNet2018}, which captures complex features with fewer parameters by expanding and then reducing the feature number. Additionally, Fused-MBConv blocks, introduced for that model, optimizes MBConv by merging pointwise and depthwise convolutions into a single step. 
These innovations contribute to the model's parameter efficiency and performance.
The EfficientNetV2 also appears more adapted to our problem than state-of-the-art Vision Transformers (ViTs) \cite{ViT2020}, such as Swin \cite{Swin2021} and SwinV2 \cite{SwinV22021}, as they only provide slightly better performance, while being harder to fine-tune with small datasets. 
Indeed, ViTs generally require larger datasets, extensive data augmentation, and regularization techniques \cite{Steiner2021}, which can also reduce the effectiveness of transfer learning \cite{Kornblith2019}.
We choose the EfficientNetV2-S variant with 20 million parameters as it offers a good compromise between performance and ease of fine-tuning for our relatively small training dataset. 
We remove the top convolutional and prediction layers and replace them with a final convolutional layer with 16 filters, followed by batch normalization and Swish activation. 
This configuration is more suited to our binary classification problem and smaller dataset than the original 1280 filters. 
This final convolution block is followed by a global pooling layer, resulting in 16 features that are fed to a final dense layer with 2 neurons and a softmax activation, outputting the probabilities for the negative and positive classes. 
%This design allows for more comprehensive explainability and facilitates future extension to multi-class classification.
\newline
\newline
% \textit{Single inputs}
% \newline
% \newline
To adapt grayscale images to the model, which is pre-trained on RGB images, we duplicate each grayscale image twice to create a 3-channel input. 
Each resulting model will be denoted by \textit{Efn-w}, where \textit{w} is the corresponding wavelength in Ångstroms, and \textit{Efn-$B_{LOS}$} for the magnetogram.
\newline
\newline
% \textit{Multimodal inputs}
% \newline
% \newline
For single-timestep multimodal inputs, we compare the following approaches:

\begin{enumerate}
\item \textit{Pretrained EfficientNetV2-S on RGB combinations of single inputs}. This approach leverages the transfer learning capabilities of the model originally trained on RGB images. We compare two combinations: a full coronal combination of 193 Å, 211 Å, and 94 Å wavelengths (\textit{Efn-193x211x94}), and a cross-atmospheric combination of the magnetogram, 304 Å, and 94 Å wavelengths (\textit{Efn-$B_{LOS}$x304x94}).
\item \textit{Features fusion from single-wavelength models}. 
In this approach, we combine features extracted from separately trained single-input models into a unified fully connected layer to investigate the linear complementarity between features from different inputs. Specifically, we perform this fusion for the cross-atmospheric combination of the magnetogram, 304 Å, and 94 Å wavelengths. The resulting model is denoted \textit{EfnFuse-$B_{LOS}$x304x94}.
\end{enumerate}

\subsubsection{Video Models}

\begin{figure*}[h!]
    \centering
    \textbf{VideoLENS Diagram}\par\medskip
    \includegraphics[width=0.99\textwidth]{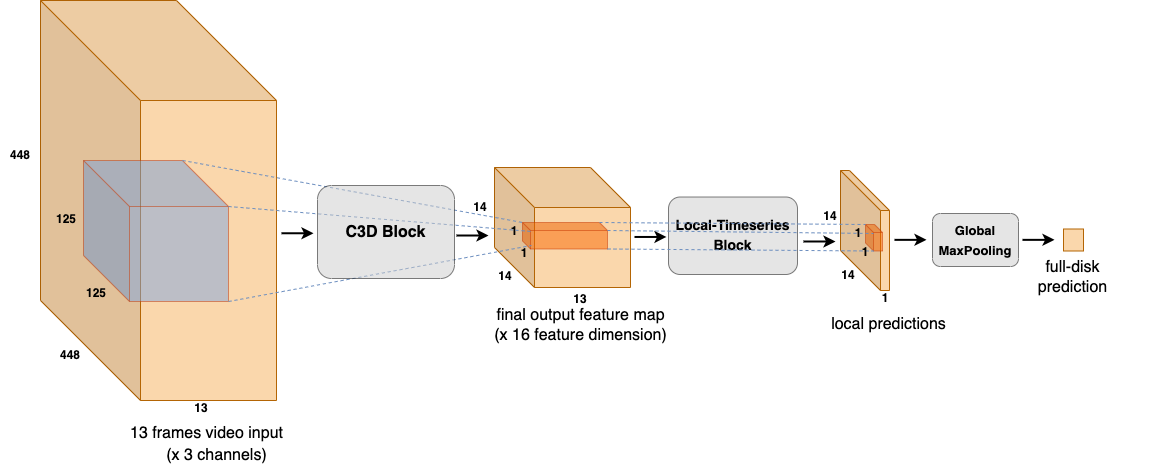}
    \caption{
    Video Local Event Neural System (VideoLENS) Diagram. 
    The architectures of the C3D and Local-Timeseries blocks are detailed in Table \ref{tab:SolVideoLENS}. 
    The blue initial block represents the original receptive field of the time series shown in dark orange. 
    The final local predictions are derived from these features, providing event predictions that are localized to their respective original receptive fields.
    \vspace{1.7ex}
    }%\ref{alg:cv_balance}
    \label{fig:figVLENS}
\end{figure*}

To incorporate temporal dynamics in analyzing the best-performing inputs, we study models using videos composed of 13 frames spaced 2 hours apart, covering the 24 hours preceding the forecasting window. Previous studies, such as \cite{Guastavino-video, guastavino2023operational}, have performed flare forecasts on magnetogram videos of \ac{AR}s using combinations of 2D-CNN and \ac{LSTM}, while \cite{Sun2022Video3D} utilized 3D-Convolutions. 
Our study, however, focuses on full-disk videos rather than \ac{AR}s and compares EUV multimodal inputs to magnetograms alone.
Full-disk videos present unique challenges, and we propose a novel deep learning model designed to effectively infer local solar events from such videos. 
While 2D convolutions and associated downsampling techniques—such as strides and max pooling—emphasize frame-dominant features, these features may originate from different \ac{AR}s across distinct full-disk frames. 
This can complicate the learning of meaningful temporal patterns when using subsequent timeseries models like \ac{LSTM}s, as the resulting time series of features may correspond to different \ac{AR}s at different timesteps.
In contrast, 3D convolutions better preserve the temporal coherence of features across frames.
We propose a hybrid approach, \ac{VideoLENS}, for full-disk videos that maintains this temporal coherence and captures short-term patterns using 3D convolutions, while also leveraging long-term dependencies through \ac{LSTM} cells and an attention mechanism (\cite{Attention2017}).
The \ac{VideoLENS} architecture is illustrated in Figure \ref{fig:figVLENS} and detailed in Table \ref{tab:SolVideoLENS}. 
The model begins with an initial block of 3D convolutions (\textit{C3D Block}) that scales the original full-disk input down to features with a spatial scale comparable to \ac{AR}s. 
Specifically, the \textit{C3D Block} is designed so that each pixel in the final output feature map has an original spatial receptive field slightly larger than the largest possible \ac{AR}s, accounting for their rotation during the 24 hours covered by the 13 input frames.
As shown in Figure \ref{fig:figVLENS}, each final time series of features (indicated in dark orange) consequently contains information localized to the corresponding original receptive field (shown in blue), ensuring that the time series accurately represents the temporal evolution of the corresponding solar region.
A \textit{Local-Timeseries Block} is then applied to each of these feature timeseries to derive more complex temporal features then used for local predictions. 
This block starts with a multi-head attention layer that computes attention scores along the temporal dimension, ensuring that each timestep is contextually aware of the entire time series and emphasizing the most relevant parts. 
An \ac{LSTM} layer is then used to learn long- and short-term features localized to \ac{AR}s. 
Subsequently, a pixel-local prediction layer, consisting of a fully 
connected neuron with sigmoid activation, predicts the probability of a flare at each spatial region.
The final full-disk prediction is derived as the maximum of these local predictions, obtained through a GlobalMaxPooling layer that aggregates the local predictions.
The resulting model can therefore aslo be used as a semi-supervised framework able to learn spatial label while only receiving non spatial one at training.
It thus extend the \ac{P-CNN} presented in \cite{Francisco2024} to video input, while removing the limitations of the patches artificial boundaries of this former model.
Similarly to the \ac{P-CNN}, precise position estimation of the predicted events could then be calculated from gradient methods for each of the local prediction.

\begin{table*}[h]
\caption{VideoLENS Architecture} % title of Table
\centering % used for centering table
\begin{tabular}{l r c c c c} % centered columns (4 columns)
\hline\hline %inserts double horizontal lines
Layer &  Kernel & Stride & Regularisation & Activation & Output \\ [0.5ex] % inserts table
%heading
\hline 
\\ [-1.5ex]% inserts single horizontal line
\textit{C3D block} & &  & & & (13, 14, 14, 16) \\ [0.1ex]
\hline %inserts single line
\\ [-2ex]
Convolution3D (1) & 16 x (3, 3, 3) & (1, 2, 2) & BN & Swish & (13, 224, 224, 16) \\ [0.7ex]
MaxPooling3D & (3, 3, 3)  & (1, 2, 2) & \o & \o & (13, 112, 112, 16) \\ [0.7ex]
Convolution3D (2) & 16 x (3, 3, 3) & (1, 2, 2) & BN & Swish & (13, 56, 56, 16) \\ [0.7ex]
Convolution3D (3) & 16 x (3, 3, 3) & (1, 1, 1) & BN & Swish & (13, 56, 56, 16) \\ [0.7ex]

Convolution3D (4) & 8 x (3, 3, 3) & (1, 1, 1) & BN & Swish & (13, 56, 56, 8) \\ [0.7ex]
Convolution3D (5) & 8 x (3, 3, 3) & (1, 1, 1) & BN & Swish & (13, 56, 56, 8) \\ [0.7ex]

Convolution3D (6) & 8 x (3, 3, 3) & (1, 2, 2) & BN & Swish & (13, 28, 28, 8) \\ [0.7ex]
Convolution3D (7) & 8 x (3, 3, 3) & (1, 1, 1) & BN & Swish & (13, 28, 28, 8) \\ [0.7ex]

Convolution3D (8) & 8 x (3, 3, 3) & (1, 1, 1) & BN & Swish & (13, 28, 28, 8) \\ [0.7ex]
Convolution3D (9) & 8 x (3, 3, 3) & (1, 1, 1) & BN & Swish & (13, 28, 28, 8) \\ [0.7ex]
Convolution3D (10) & 16 x (3, 3, 3) & (1, 2, 2) & BN + DP(0.2) & Swish & (13, 14, 14, 16) \\ [0.7ex]
\hline %inserts single line
\\ [-1.5ex]
\textit{Local Timeseries Block} & &  & & & (14, 14, 1) \\ [0.1ex]
\hline %inserts single line
\\ [-2ex]
%Dropout(0.2) & \o & \o & \o & \o & (13, 16) \\ [0.7ex]
MultiHead-Attention & 4-head x 16-dim & residual connection & LN + DP(0.2) & \o & (13, 16) \\ [1ex]
%Dropout(0.2) & \o & \o & \o & \o & (13, 16) \\ [0.7ex]
LSTM & 16-cells & \o & BN & \o & (16) \\ [0.7ex]
Dense (local predication) & 1 & \o & \o & Sigmoïd & (1) \\ [0.2ex]
\hline \\ [-1.7ex]
GlobalMaxPooling & \o  & \o & \o & \o & (1) \\ [0.7ex]

% CrossAtmo3D & $3E-4$ & $1E-4$ & na & na  \\ [1ex]
% [1ex] adds vertical space
\hline %inserts single line
\end{tabular}
\\ [1ex]
\parbox{18cm}{Architecture of the \ac{VideoLENS}. BN : Batch Normalisation. LN : Layer Normalisation. DP(0.2) : dropout with a rate of 0.2. The \textit{Local Timeseries Block} is applied to each spatial-point of the attention layer output, i.e. to ervery (13, i, j, 16) with $(i,j)\in \llbracket 1,14 \rrbracket^{2}$
\vspace{2ex}
}
\label{tab:SolVideoLENS}. % is used to refer this table in the text

\end{table*}

\subsection{Training and hyperparameters}

We use the same full disk \ac{CV} method as presented in \cite{Francisco2024}.
Specifically temporal buffers of 27-days are used to create independent temporal chunks that are selected by optimisation to build balanced training and validation folds.
This result in a 5-fold  \ac{CV} done on the period ranging from May 2010 to December 2019, while the period from January 2020 to April 2023 is unaltered, meaning every sample and the natural distribution of the data is preserved and kept as a complementary test set.
Our models are trained using the Adam optimizer \cite{Kingma2014}. 
We conducted a Bayesian parameter search for the EfficientNetV2 model with magnetogram inputs (\textit{Efn-$B_{LOS}$}) to determine an optimal learning rate and the potential inclusion of a weight decay parameter. 
Although decoupled weight decay regularization \cite{Loshchilov2017} is typically employed to reduce overfitting risk, we found it to have minimal impact on the fine-tuning of the pre-trained EfficientNetV2. 
This observation is consistent with the findings of \cite{Kornblith2019}, which indicated that regularization methods may not always be useful in various transfer learning scenarios.
Our results also showed that fine-tuning the EfficientNetV2 generally converges in fewer than 10 epochs across most parameter combinations tested. This rapid convergence aligns with \cite{Kornblith2019}, who reported that fine-tuning pre-trained models typically requires significantly fewer epochs than training from scratch —approximately 17 times faster.
Based on the parameter search results, we train all EfficientNetV2-based models with a learning rate of $3 \times 10^{-4}$ for 15 epochs and  without weight decay regularisation.
For the \ac{VideoLENS} models, a manual parameter search revealed that a slight weight decay regularization of $1 \times 10^{-5}$ marginally enhances performance. 
We also found an optimal learning rate that remains the same as for the EfficientNetV2 models, at $3 \times 10^{-4}$.
Finally, we use weighted binary cross-entropy as the loss function to fully address class imbalance.

\subsection{Evalutation}

To compare the predictive power of the models, we focus on the \ac{AUC} of the \ac{ROC}, the \ac{TSS}, the \ac{HSS} and the \ac{MCC}.
% \subsubsection{ROC AUC}
The \ac{ROC} curve gives a model's achievable recall as a function of the \ac{FPR} over the different decision thresholds.
It directly relates to the \ac{TSS} which is one of the preferred metrics in flare forecasting studies (\cite{Bloomfield2012}), as the vertical distance between the diagonal and the \ac{ROC} correspond to the \ac{TSS} at the corresponding threshold.
Optimal \ac{ROC} \ac{AUC} values therefore improve the likelihood of finding a threshold with good \ac{TSS} values.
The \ac{ROC} \ac{AUC} is an easily interpretable value of a classifier's discriminatory power as it corresponds to the probability that among two random positive and a negative samples, the classifier will assign a  higher probability output to the positive one.
% \subsubsection{TSS}
The \ac{TSS}, (\cite{TssPeirce}, \cite{TssHanssen}, \cite{Woodcock1976}) is defined as the difference between the recall and the \ac{FPR}.
Also known as the (bookmaker) informedness, Peirce's index or Younden's J index, it is also equal to the balanced accuracy rescaled between -1 and 1 :

% \newpage
% \textcolor{white}{.}
% \newpage
% \textcolor{white}{.}
% \newpage

\begin{equation}\label{eq:tss}
TSS = \frac{TP}{TP + FN} - \frac{FP}{FP + TN} = TPR - FPR,
\end{equation}
or equivalently
\begin{equation}
  TSS =   TPR + TNR - 1,
\end{equation}

where TPR and TNR represent the true positive rate and true negative rate, respectively.
% \subsection{Limitations for imbalanced problems}
It is noted that the \ac{TSS} and \ac{ROC} \ac{AUC} suffer some limitations in measuring a model's ability to discriminate properly between two classes.

\newpage
\textcolor{white}{.}
\newpage
\textcolor{white}{.}
\newpage

The two metrics only encompass information about the success rate of detection in each class, without information about the precision with which predictions are made.
In imbalanced cases, this flaw allows high \ac{TSS} and \ac{ROC} \ac{AUC} to mask models that distinguish between negative and positive with poor precisions (\cite{Jeni2013}).
In the specific case of flare forecasting, some studies, exhibit \ac{TSS} and \ac{ROC} \ac{AUC} that are higher for M+ forecasting than C+ forecasting on the same datasets (e.g. \cite{Guastavino-video}, \cite{Leka2023}).
This gives the misleading impression that the models discriminate better stronger flare than medium ones, whereas they usually have precision at least twice as small when predicting these stronger events.
% \subsubsection{HSS}
The \ac{HSS}, defined in equation \ref{eq:hss}, was originally introduced by \cite{HssHeidke} for evaluating weather forecasts. 
Also known as Cohen's Kappa index, it is widely used in flare forecasting to compare a model's skill relative to a random guess model (\cite{Camporeale2019}). 
The \ac{HSS} is particularly comprehensive for imbalanced datasets, as it combines information from the TSS with the Markedness, which synthesizes the class-wise precision of the model’s predictions for each class, including both Positive Predictive Value (PPV) and Negative Predictive Value (NPV). 
The Markdness is the precision analog to how the TSS captures class-wise accuracy, synthesizing True Positive Rate (TPR) and True Negative Rate (TNR).
For a deeper understanding of how the \ac{HSS} integrates both TSS and Markdness, one can refer to the mathematical analysis presented in \cite{mccVsHss}. 
This analysis highlights the close relationship between the \ac{HSS} and the binary correlation coefficient, which serves as a geometric average between TSS and Markdness. 
Specifically, from the $\alpha$ and $\beta$ coefficients in \cite{mccVsHss}, the \ac{HSS} can be understood as the harmonic mean of Informedness (TSS) and Markedness, adjusted for the model's frequency bias.

\begin{equation}\label{eq:hss}
HSS = 2*\frac{TP*TN - FN*FP}{P(TN+FN) + N(TP+FP)},
\end{equation}

where $P$ and $N$ represents the total number of positive and negative samples, respectively. 
% \subsubsection{MCC}
The \ac{MCC} (\cite{MccMatthews}), defined in equation \ref{eq:mcc}, is the Pearson correlation coefficient between two binary variables.
It contains similar information to the \ac{HSS} with constant even weighting between TSS and the Markdness.
It thus provides an agnostic measure of a model discriminatory power as it takes into account all the basic confusion matrix rate of a model giving equal importance to each of them.
This makes it particularly suitable to assess models discriminatory powers on imbalanced problems.
We provide a more detail comparison of the \ac{TSS}, \ac{HSS} and \ac{MCC} and discussion of their strength and limitations in \cite{Francisco2024} (section 2.1.1 of and associated appendix).

\begin{equation}\label{eq:mcc}
MCC = \frac{TP*TN - FN*FP}{\sqrt{P(TN+FN)*N*(TP+FP)}}
\end{equation}

% \newpage
\begin{figure*}[h!]
    \centering
    \textbf{ROC AUC}\par\medskip
    \includegraphics[width=0.65\textwidth]{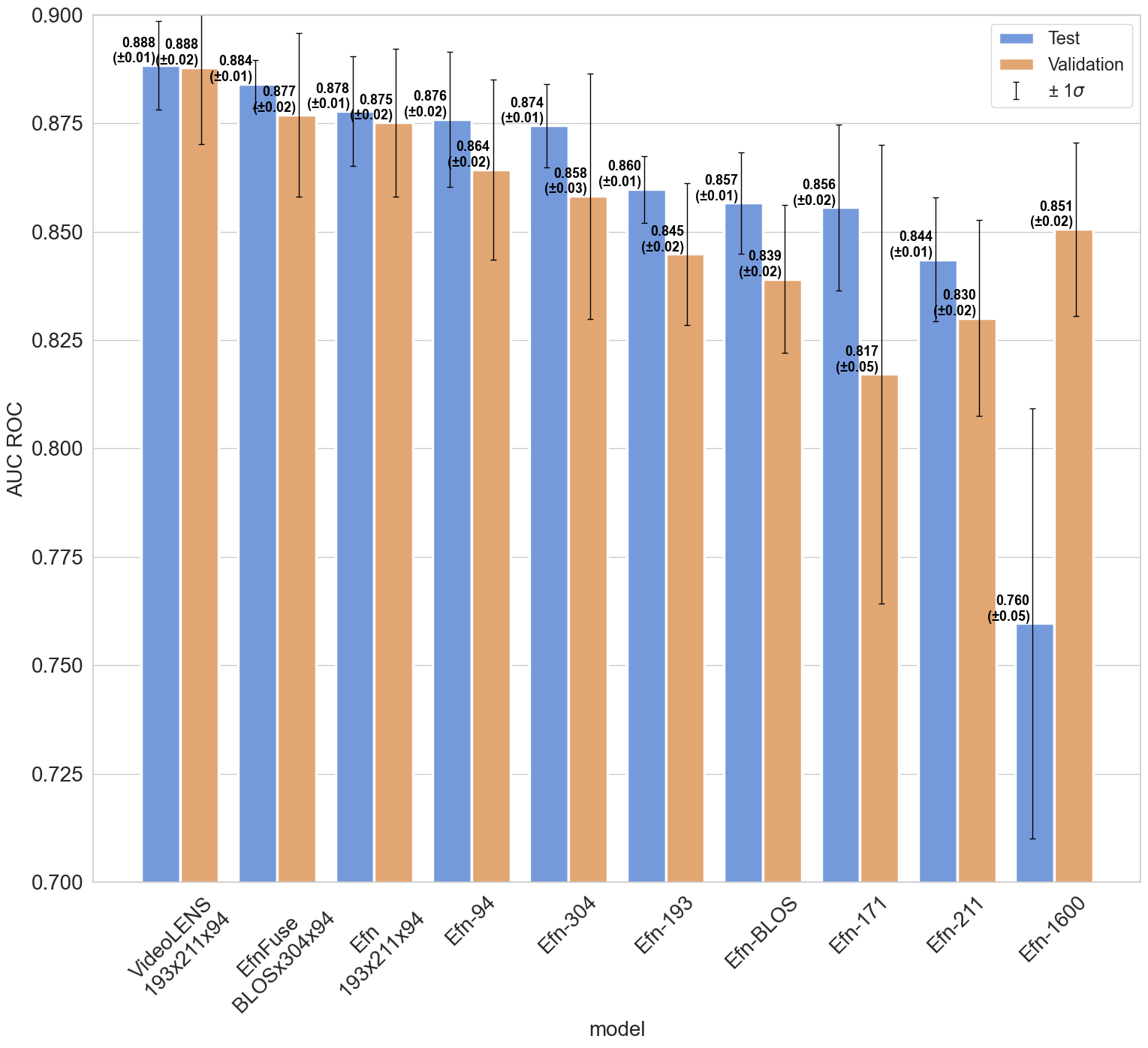}
    \caption{ROC AUC of various models. Efn-\textit{input} denotes the EfficientNetV2-S model trained on individual frame \textit{input}. 
    EfnFuse-\textit{inputs} refers to a logistic regression model utilizing features extracted by EfficientNetV2-S models trained on each of the distinct \textit{inputs}.
    VideoLENS represents the video-based models.
    }
    \label{fig:figAUC}
\end{figure*}

\begin{figure*}[h!]
    \centering
    \textbf{TSS}\par\medskip
    \includegraphics[width=0.65\textwidth]{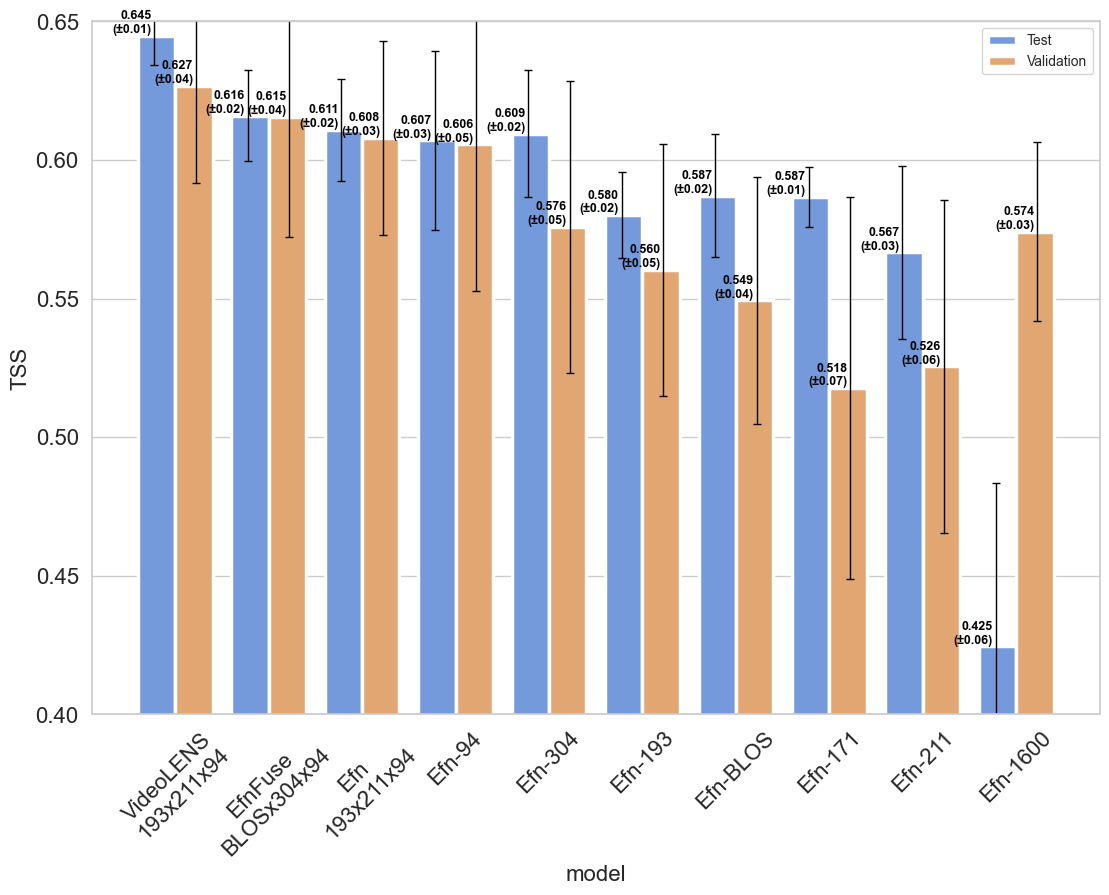}
    \caption{Models' TSS.
    Models' labels are the same  as the ones described in \ref{fig:figAUC}.
    }
    \label{fig:figTss}
\end{figure*}

% \newpage

\begin{figure*}[h!]
    \centering
    \textbf{HSS}\par\medskip
    \includegraphics[width=0.68\textwidth]{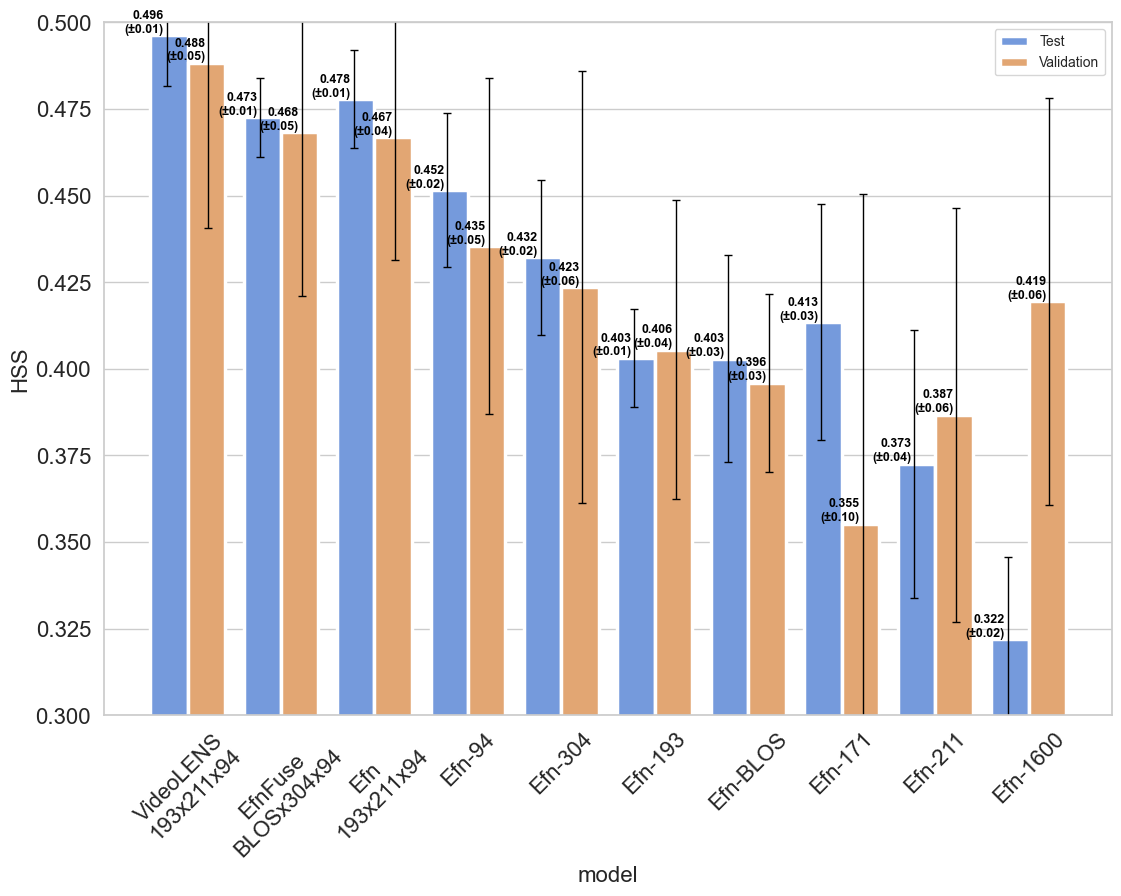}
    \caption{Models' HSS.
    Models' labels are the same  as the ones described in \ref{fig:figAUC}.
    }
    \label{fig:figHss}
\end{figure*}

\begin{figure*}[h!]
    \centering
    \textbf{MCC}\par\medskip
    \includegraphics[width=0.68\textwidth]{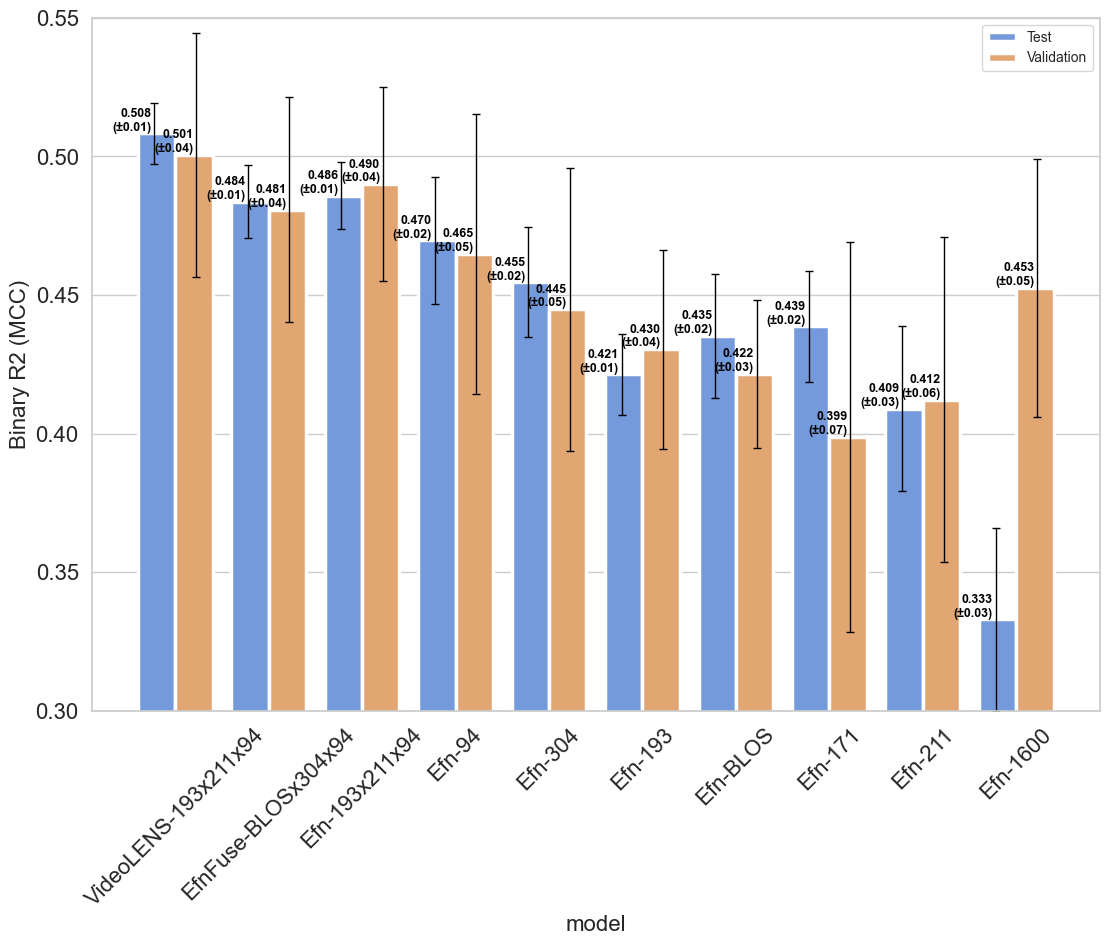}
    \caption{Models' MCC.
    Models' labels are the same  as the ones described in \ref{fig:figAUC}.
    }
    \label{fig:figMcc}
\end{figure*}

% \newpage

\section{Results \& Discussions}

Figures \ref{fig:figAUC}, \ref{fig:figHss}, \ref{fig:figTss} and \ref{fig:figMcc} present the performance metrics of the models: \ac{ROC} \ac{AUC}, \ac{HSS}, \ac{TSS} and \ac{MCC} respectively.
The plots rank the models based on their performance, with a consistent ranking observed across all three metrics. 
Notably, while \ac{ROC} \ac{AUC} values are clustered within a narrow range (+16\% from the worst to the best model), the \ac{MCC}, \ac{HSS} and \ac{TSS} demonstrate more pronounced disparities in model performance (with maximum performance gaps of +54\% and +51\%, respectively).
The results also show stability between validation and test sets, indicating good generalization capabilities. 
However, the model using the 1600\AA \ wavelength exhibited a marked decline in \ac{HSS} and \ac{TSS} by 24\% and 26\%, respectively, on the test set compared to the validation set. 
This performance drop may be attributed to the learning of unrelevant features from the 1600\AA \ wavelength that persist across the 27-day buffer period of the \ac{CV} folds, leading to potential overfitting on the validation data. 
In addition, there might be some pattern specific to the solar cycle's 24th period which does not generalize to the next cycle.
This finding underscore that even rigorous \ac{CV} methods do not guarantee performance generalization in operational settings. 
Consequently, incorporating a separate operational test set, as advocated by \cite{Cinto2020}, appears essential for a more comprehensive evaluation of model performance.

\subsection{Coronal and Chromospheric Features are Most Predictive of Flares}

The 94\AA\ wavelength provides the most discriminative features for flare prediction, consistent with findings from \cite{Leka2023} and \cite{Sun2023}. 
Compared to the line-of-sight magnetograms, the \textit{Efn-94} model shows a significant improvement, with a 3.4\% increase in \ac{TSS} on the test set and a 10.4\% increase on the validation set. Additionally, \ac{HSS} values improve by 12\% on the test set and 9\% on the validation set.
Although the 193\AA\ and 211\AA\ wavelengths demonstrate similar or lower predictive performance to the line-of-sight magnetograms, their combination with the 94\AA\ wavelength in the \textit{Efn-193x211x94} model results in a further enhanced \ac{HSS}, showing an overall increase of 18\% on both the test and validation sets compared to the magnetograms alone. 
This suggests that the relative intensity variations between these three wavelengths may serve as effective precursors for flare events.
Chromospheric observations also contribute valuable predictive features.
While the lower chromospheric wavelength of 1600\AA\ exhibits weak generalization performance on the test set, the upper chromospheric wavelength of 304\AA\ emerges as the second most predictive single input, following the 94\AA\ wavelength.

% \newpage
% \textcolor{white}{.}
% \newpage
\subsection{Features from Different Atmospheric Layers Complement Each Other}

Combining features from various solar atmospheric layers enhances overall performance compared to using the single best-performing layer. 
The model \textit{EfnFuse-$B_{LOS}x304x94$}, which integrates features from $Efn-B_{LOS}$, $Efn-304$, and $Efn-94$, outperforms the $Efn-94$ model by 5.7\% and 7.4\% in \ac{HSS} on the test and validation sets, respectively. 
While improvements in \ac{TSS} and \ac{ROC} \ac{AUC} are more modest, the notable increase in \ac{HSS} indicates enhanced precision, with improved performances in discriminating ambiguous cases.
% \newpage
% \textcolor{white}{.}
% \newpage
We found that, cross-atmospheric feature complementary is most effective through feature fusion after training individual models for each input (\textit{EfnFuse-$B_{LOS}x304x94$}). Training a single model directly on the cross-atmospheric channel combination with the pretrained EfficientNetV2 ((\textit{Efn-$B_{LOS}x304x94$}) does not surpass the performance of the model using the most predictive channel alone, conversely to the coronal wavelength combination (\textit{Efn-$193x211x94$}). This result might arises from the 2D-CNN architecture, where combining multiple channels in early convolutional layers treats them as a distinct feature dimension.
\newpage
\textcolor{white}{.}
\newpage
Filters learn different weights for each channel, and sum their outputs to form the feature map.
This approach is particularly effective for features that complement each other linearly, such as the coronal wavelength channels in the \textit{Efn-$193x211x94$}, which capture different emission lines of similar coronal structures. 
In contrast, magnetograms, chromospheric images, and coronal images — representing more distinct physical layers and processes - might not combine linearly as well in the first convolutional layers.
Alternative approaches to improve cross-atmospheric models performance could include using 3D convolutions in the initial layer of a single model, or training on all layers simultaneously the distinct bock of the \textit{EfnFuse-$B_{LOS}x304x94$}, rather than pre-training them separately. 
However, this latter method may require larger datasets or data augmentation to prevent overfitting due to the increased model complexity.
Ultimately, while combining features from different atmospheric layers enhances performance, it does not significantly surpass the results achieved with the three coronal channels in the  \textit{Efn-$193x211x94$} model.

\subsection{Temporal Dynamics Enhance Predictive Performance Compared to Static Features}

The \textit{VideoLENS-193x211x94} model outperforms all others, showing up to a 5.8\% improvement in \ac{ROC} \ac{AUC}, a 14\% increase in \ac{TSS}, and a 23\% enhancement in \ac{HSS} over the \textit{Efn-$B_{LOS}$} model. 
When compared to the static single-frame model \textit{Efn-193x211x94}, the \textit{VideoLENS-193x211x94} achieves improvements of up to 5.6\% in \ac{TSS} and 4.4\% in \ac{HSS}.
Attempts to combine other wavelength pairs, such as the cross-atmospheric combination $B_{LOS}x304x94$, using both parallel processing and an RGB-like processing approach similar to \textit{VideoLENS-193x211x94}, did not yield models significantly surpassing the performance of the \textit{Efn-193x211x94} single-frame coronal model.
These results underscore that the $193x211x94$ wavelength combination provides some of the most predictive information for flare forecasting among the atmospheric layer combinations tested.
The \textit{Efn-193x211x94} and \textit{VideoLENS-193x211x94} models achieve \ac{HSS} scores of 0.46 and 0.48, respectively, significantly surpassing other known full-disk models for forecasting M+ flares within 24 hours. 
For context, \cite{Pandey2023pub} reports an \ac{HSS} of 0.35 for a similar problem using full-disk line-of-sight magnetograms over a comparable validation period.

% \subsection{VideoLENS ablation study}

% The VideoLENS architecture was found to be the best performing against a significant number of variations.
% Both the local-wise predictions and the time-series block slightly improve the AUc ROC over c3Ds with global pooling before the prediction layer.
% Smaller, or bigger models (more filters, more - or - bigger attention and LSTM layers) did not improve performances, probably due to the limited size of the training set.
% At an equal number of parameters, models with fewer C3D layers made of more filters were leading to poorer results.
% Finally, the C3D alone appeared significantly better than a 2D-CNN-LSTM.
% This confirms that the 2D-CNN-LSTM approach is less suited for the full-disk approach as the CNN  may converge on features from distinct AR across the different frames in the case of multiple active AR over the video period.

\subsection{VideoLENS Ablation Study}

The VideoLENS architecture was evaluated through an ablation study to determine the impact of various components on model performance. 
This study revealed that incorporating local-wise predictions and the time-series block offered slight improvements in \ac{ROC} \ac{AUC} compared to models using only C3D with global pooling.
Increase in model size, through the number of filters or parameters of the attention and LSTM layers, did not result in significant performance gains, likely due to the modest size of the training dataset.
At equal number of parameters, fewer C3D layers but more filters performed less effectively, indicating that simply increasing the number of filters is not beneficial without careful architectural tuning. 
More importantly, every C3D-based models consistently outperformed 2D-CNN-LSTM architectures, highlighting that the latter may not be as effective for analyzing full-disk videos. 
This  is attributed to the tendency of the 2D-CNN component to capture features from various ARs across different frames, when multiple ARs coexist during the video.

\subsection{The last and first flare challenge}

An examination of model performance revealed that all models exhibited roughly null TSS and HSS scores when evaluated on the samples' subsets where the label of the predicted time window differed from the previous time window. 
This confirm that our previous observations from \cite{Francisco2024} generalise to a wide range a input modalities and to the use of low temporal resolution data to forecast flares.
Specifically, the models struggle to forecast events that occurred in the next 24 hours following a period of inactivity, as well as to forecast a quiet period when flares occurred in the past 24 hours.
This indicates that while the models can effectively predict flares when there is a persistent level of activity, they face substantial difficulties in scenarios involving activity changes, with performance akin to random guessing in these transitional cases.
The current most predictive flare precursors might, therefore, be more indicative of the current activity level of an \ac{AR} rather than being actually predictive.

\newpage
\section{Conclusion and Future Works}
For years, studies on solar activity forecasting have been employing the photospheric AR complexity (computed either from continuum images or from magnetic field information) to predict the AR flare activity \cite[e.g.,][]{schrijver2007characteristic, korsos2015flare, benvenuto2018hybrid, korsos2020solar, Cicogna_2021}.
However, it is widely accepted that the coronal magnetic topology and dynamics are crucial to driving and then triggering the release of solar eruptions.\\
Since direct information on the coronal magnetic field is still not available for such studies, researchers are using as a proxy the EUV multiwavelength images to reveal the onset mechanism of solar flares in case studies \cite[][]{Gosain_2012, Imada_2014, Bamba_2014} and also for statistic approaches \cite[][]{Dissauer2023, Leka2023}.
Such uses have been recently extended to deep-learning approaches to automatically extract flare precursors from EUV images to build models for flare forecasting \cite[e.g.,][]{Sun2023, Nishizuka2020}.
Following this trend, we created a suitable dataset for DL  approaches and investigated different 2D and 3D CNN models, exploiting the multiwavelength EUV images from the AIA instrument, the photospheric LoS magnetogram from the HMI instrument, and the temporal dimension.\\
Below is a summary of the main findings from this study:
\begin{enumerate}
    \item \textbf{Enhanced Forecasting with Coronal Wavelengths:} 
    The use of coronal EUV wavelengths emissions notably enhances forecasting performance compared to relying on photospheric line-of-sight magnetograms. 
    The combination of emission lines at 193 \AA, 211 \AA, and 94 \AA \ proves particularly effective. 
    This suggests that the relative intensity between EUV wavelengths may provide crucial information about flare precursors. 
    Future research will focus on a detailed investigation of the \textit{Efn-193x211x94} model to identify specific precursors through explainability methods and gain further insights into flare mechanisms.
    \newline
    \item \textbf{VideoLENS Architecture and Temporal Dynamics:} 
    The VideoLENS architecture is an efficient way to forecast localized solar events from full-disk videos. 
    Notably, the use of such model to incorporate temporal dynamics in flare predictive features yields superior performance compared to single-timestep input models. 
    Specifically, the VideoLENS model using the 193 \AA, 211 \AA, and 94 \AA \ wavelengths achieves an HSS of 0.50 and a TSS of 0.65, substantially outperforming typical full-disk models that forecast M-class flares within the next 24 hours.
    \newline
    \item \textbf{Challenges in Forecasting Activity Changes:} All tested models exhibit limitations in forecasting transitions in activity levels.
    This suggests that the features derived from the studied inputs are more indicative of current activity levels rather than exclusive predictors of upcoming flares. 
    While the inherent stochasticity of complex physical systems may partly explain these limitations in forecasting activity changes, further research could be of interest. 
\end{enumerate}
Future work will explore higher temporal resolution features than the 2 hours used in this work as well as different forecasting window sizes, and the addition of other data modalities such as \ac{SXR} and \ac{EUV} flux timeseries.

\newpage
\begingroup
% \Large     
\begin{acknowledgements}

This research is part of the SWATNet project which is funded by the European Union’s Horizon 2020 research and innovation program under the Marie Sklodowska-Curie Grant Agreement No 955620. 

This research has also been carried out in the framework of the CAESAR project, supported by the Italian Space Agency and the National Institute of Astrophysics through the ASI-INAF n.2020-35-HH.0 agreement for the development of the ASPIS prototype of the scientific data centre for Space Weather. 

This study was also produced within the IA and the CITEUC.
IA is supported by Fundação para a Ciência e a Tecnologia (FCT, Portugal) through the research grants UIDB/04434/2020 and UIDP/04434/2020.
CITEUC is funded by National Funds through FCT - project UIDP+UIDB/00611/2019.

Project partially funded under the National Recovery and Resilience Plan (PNRR), Missione 4 “Istruzione e Ricerca” – Componente C2 – Investimento 1.1, “Fondo per il Programma Nazionale di Ricerca e Progetti di Rilevante Interesse Nazionale (PRIN)” – Call for tender No. 1409 of 14/09/2022 of Italian Ministry of University and Research funded by the European Union – NextGenerationEU Award Number: P2022RKXH9, Concession Decree No. 1397 of 06/09/2023 adopted by the Italian Ministry of University and Research, project CORonal mass ejection, solar eNERgetic particle and flare forecaSTing from phOtospheric sigNaturEs (CORNERSTONE).

\end{acknowledgements}
\endgroup

\newpage
% \textcolor{white}{.}
% \newpage
\bibliographystyle{aa} 
\bibliography{references.bib} 

\addcontentsline{toc}{chapter}{List of Acronyms}
\section{List of Acronyms}%\thispagestyle{fancy}
\begin{acronym}[SWATNet Project]\itemsep2pt
\acro{1D}{1-Dimensional}
\acro{2D}{2-Dimensional}
\acro{3D}{3-Dimensional}
\acro{AA}{Academy of Athens}
\acro{ASRO}{Aboa Space Research Oy}
\acro{AI}{Artificial Intelligence}
\acro{AR}{Active Region}
\acro{ASCII}{American Standard Code for Information Interchange}
\acro{AU}{Astronomical Unit}
\acro{CAM}{Class Activation Maps}
\acro{CDAWeb}{Coordinated Data Analysis Web}
\acro{CDF}{Common Data Format}
\acro{CDP}{Career Development Plan}
\acro{CET}{Central European Times}
\acro{CME}{Coronal Mass Ejection}
\acro{CNN}{Convolutional Neural Network}
\acro{CSV}{Comma-Seperated Values}
\acro{DHO}{Debrecen Heliophysical Observatory}
\acro{DOI}{Digital Object Identifier}
\acro{EAB}{External Advisory Board}
\acro{ECAS}{European Commission Authentication System}
\acro{ECTS}{The European Credit Transfer and Accumulation System}
\acro{ELTE}{Eötvös Loránd University}
\acro{ESA}{European Space Agency}
\acro{ESR}{Early Stage Researcher}
\acro{EU}{European Union}
\acro{EUV}{Extreme Ultraviolet}
\acro{EUHFORIA}{EUropean Heliospheric FORecasting Information Asset}
\acro{EUHFORIA 2.0}{EUropean Heliospheric FORecasting Information Asset 2.0}
\acro{FARe}{False Alarm Rate}
\acro{FAR}{False Alarm Ratio}
\acro{FGE}{Fluid Gravity}
\acro{FITS}{Flexible Image Transport System}
\acro{FP7}{The Seventh Framework Programme of the European Union}
\acro{GA}{Grant Agreement}
\acro{GAN}{Generative Adversarial Network}
\acro{GCS}{Graduated Cylindrical Shell}
\acro{GSO}{Gyula Bay Zoltán Solar Observatory}
\acro{HEEQ}{HEliocentric Earth equatorial}
\acro{HMI}{Helioseismic and Magnetic Imager}
\acro{HSPF}{Hungarian Solar Physics Foundation} 
\acro{ICME}{Interplanetary Coronal Mass Ejection}
\acro{IDL}{Interactive Data Language}
\acro{I/O}{Input/Output}
\acro{ITN}{Innovative Training Network}
\acro{IMF}{Interplanetary Magnetic Field}
\acro{IP}{InterPlanetary}
\acro{IPN}{Instituto Pedro Nunes}
\acro{JSOC}{Joint Science Operations Center} 
\acro{KUL}{KU Leuven}
\acro{L1}{first Lagrangian point}
\acro{LOS}{Line-Of-Sight}
\acro{LSTM}{Long Short-Term Memory}
\acro{MSCA}{Marie-Sk{\l}odowska-Curie Action}
\acro{MFM}{magnetofrictional method}
\acro{MHD}{magnetohydrodynamics}
\acro{ML}{Machine Learning}
\acro{MLP}{Multi-Layer Perceptron}
\acro{NN}{Neural Network}
\acro{NRT}{Near-Real-Time}
\acro{OA}{Open Access}
\acro{PHI}{Polarimetric and
Helioseismic Imager}
\acro{PIL}{Polarity Inversion Line}
\acro{PCDP}{Personal Career Development Plan}
\acro{PFSS}{Potential Field Source Surface}
\acro{PI}{Principal Investigator}
\acro{PSP}{Parker Solar Probe}
\acro{PTECH}{Present Technologies, LDA}
\acro{RGB}{Red-Green-Blue}
\acro{RNN}{Recurrent Neural Network}
\acro{SAS}{Space Applications Services}
\acro{SB}{Supervisory Board}
\acro{SDO}{Solar Dynamic Observatory}
\acro{SEP}{Solar Energetic Particle}
\acro{SF}{Solution Focus}
\acro{SOHO}{Solar and Heliospheric Observatory}
\acro{SolO}{Solar Orbiter}
\acro{SOLPACS}{SOLar Particle Acceleration in Coronal Shocks}
\acro{SSC}{Sheffield Solar Catalogue}
\acro{STEM}{Science, Technology, Engineering and Mathematics}
\acro{STEREO}{Solar Terrestrial Relations Observatory}
\acro{SVM}{Support Vector Machine}
\acro{SWATNet}{Space Weather Awareness Training Network}
\acro{TSS}{True Skill Statistic}
\acro{WP}{Work Package}
\acro{UC}{University of Coimbra}
\acro{UH}{University of Helsinki}
\acro{UMCS}{Maria Curie-Skłodowska University}
\acro{UNITOV}{Università degli Studi di Roma Tor Vergata}
\acro{UoI}{University of Ioannina}
\acro{USFD}{University of Sheffield}
\acro{UTU}{University of Turku}
\acro{UV}{Ultraviolet}
\acro{PRSS}{Persistent Relative Skill Score}
\acro{FSS}{F1-Skill-Score}
\acro{PR-F1}{Persistent-Relative-F1}
\acro{HSS}{Heidke Skill Score}
% \acro{TSS}{True Skill Statistic}
\acro{MCC}{Matthews Correlation Coefficient}
\acro{SPCNN}{Solar-Patch-Distributed-CNN}
\acro{P-CNN}{Patch-Distributed-CNN}
\acro{AC}{Activity-Change}
\acro{NC}{No-Change}
\acro{TPR}{True Positive Rate}
\acro{TNR}{True Negative Rate}
\acro{PPV}{Positive Predictive Value}
\acro{NPV}{Negative Predictive Value}
\acro{TP}{True Positive}
\acro{TN}{True Negative}
\acro{FP}{False Positive}
\acro{FN}{False Negative}
\acro{AIA}{Atmospheric Imaging Assembly}
\acro{SXR}{Soft X-Rays}
\acro{MPF}{Maximum Peak Flux}
\acro{CV}{Cross-Validation}
\acro{NFB}{Negative-Frequency-Bias}
\acro{VideoLENS}{Video Local Event Neural System}
\acro{AUC}{Area Under the Curve}
\acro{ROC}{Receiver Operating Characteristic}
\acro{FPR}{False Positive Rate}

\end{acronym}

\end{document}